\documentclass[conference]{IEEEtran}
\IEEEoverridecommandlockouts
\usepackage{cite}
\usepackage{algorithm}
\usepackage{algpseudocode}
\usepackage{amsmath,amssymb,amsfonts}
\usepackage{graphicx}
\usepackage{textcomp}
\usepackage{multirow}
\usepackage{booktabs}
\usepackage{soul}
\usepackage{xcolor}

\usepackage[table]{xcolor}
\usepackage{makecell}
\usepackage{hyperref}

\def\BibTeX{{\rm B\kern-.05em{\sc i\kern-.025em b}\kern-.08em
    T\kern-.1667em\lower.7ex\hbox{E}\kern-.125emX}}
    
\begin{document}

\title{LMS-AR: LMS Prediction-based Adaptive Regulator for Memory Bandwidth in Multicore Systems}

\author{\IEEEauthorblockN{Sudarshan Srinivasan}
\IEEEauthorblockA{\textit{Computer Systems Group } \\
\textit{IIIT Hyderabad}\\
Hyderabad, India \\
sudarshan.srinivasan@research.iiit.ac.in}
\and
\IEEEauthorblockN{Deepak Gangadharan}
\IEEEauthorblockA{\textit{Computer Systems Group} \\
\textit{IIIT Hyderabad}\\
Hyderabad , India \\
deepak.g@iiit.ac.in}

\and
\IEEEauthorblockN{Dip Goswami}
\IEEEauthorblockA{\textit{Electronic Systems group} \\
\textit{Eindhoven University of Technology (TU/e)}\\
Eindhoven, Netherlands \\
d.goswami@tue.nl}
}
\maketitle

\begin{abstract}
Memory bandwidth contention in multi-core systems severely impacts application performance and quality-of-service (QoS) guarantees. Regulating the shared memory bandwidth mitigates the memory performance uncertainty thereby making it a  manageable resource and improving trustworthiness of multi-core systems. 
%In a multicore system, core throttling has been a widely used technique to extenuate the impact of interference. The current solutions on MPSoC rely on features such as DRAM controller profiling and on-chip debug features to carry out the regulation by throttling. However, such features are not the norm among the hardware platforms, which limits the usefulness of these solutions on hardware platforms that do not support these features.
In this work we propose a memory bandwidth regulation mechanism \emph{LMS-AR}, i.e., LMS Prediction-based Adaptive Regulator within a Linux kernel module to distribute the memory bandwidth as a resource among the CPU cores. We describe a design in which both monitoring and  regulation is enforced from outside by a master core - which is not a dedicated controller for regulation. This allows for plugging in computationally heavy prediction and regulation algorithms  without interfering with the regulated core. An adaptive filtering technique was employed for prediction of per-core bandwidth requirement. We conducted several experiments with SPEC CPU 2017 benchmarks distributed across multiple cores. Our proposed approach demonstrated significant improvement over Memguard with respect to slowdown ratios caused due to memory contention. Our solution is hosted publicly at \href{https://github.com/ss22ongithub/LMSAdaptiveRegulator}{https://github.com/ss22ongithub/LMSAdaptiveRegulator}
\end{abstract}


\section{Introduction}
Modern embedded applications demand growing compute and memory resources, a challenge compounded as more cores contend for shared main memory. Memory-bound applications on a few cores cause contention that degrades the performance of applications on other cores. This is traditionally addressed by providing temporal isolation among competing applications.
Temporal isolation in embedded systems with shared DRAM has been studied from both hardware~\cite{Predator} and software~\cite{memguard2013} perspectives. Shared DRAM contention undermines timing predictability: when cores access memory simultaneously, they compete for banks, row buffers, and controller bandwidth, producing interference where one core's request can delay another's by orders of magnitude depending on bank conflicts, row-buffer misses, and refresh cycles.

\subsection{Motivation}
% The out-of-core approach used in \textit{Mempol} \cite{Zuepke2024} requires a separate processing element (on board micro-controller) with its own SRAM to operate. This element controls the bandwidth consumed by the other CPU cores called the regulated cores via on board debug interfaces.
% In off-the-shelf platforms, we do not have this option, and the regulator module has to execute as part of the master core (Core-0).This also causes the master core to generate its own memory transactions, which is a potential source of interference.
% In this work, we proceed with the premise that if the cores on the system are operated close to the \textit{sustainable bandwidth}\cite{ewarp} while processing the workloads in such a way that the workload requirements are satisfied, then the workload would experience only a very minimal interference.\par

The out-of-core approach in MemPol\cite{Zuepke2024} relies on a separate microcontroller with its own SRAM that regulates other cores via on-board debug interfaces. Such interfaces are  typically unavailable on off-the-shelf platforms, forcing our regulator to execute on the master core (Core-0), which then generates its own potentially interfering memory traffic. We proceed on the premise that operating cores close to the sustainable bandwidth \cite{ewarp} while still meeting workload requirements yields only minimal interference.\par

\begin{figure}[!htbp]
    \centering
    \includegraphics[width=0.5 \textwidth]{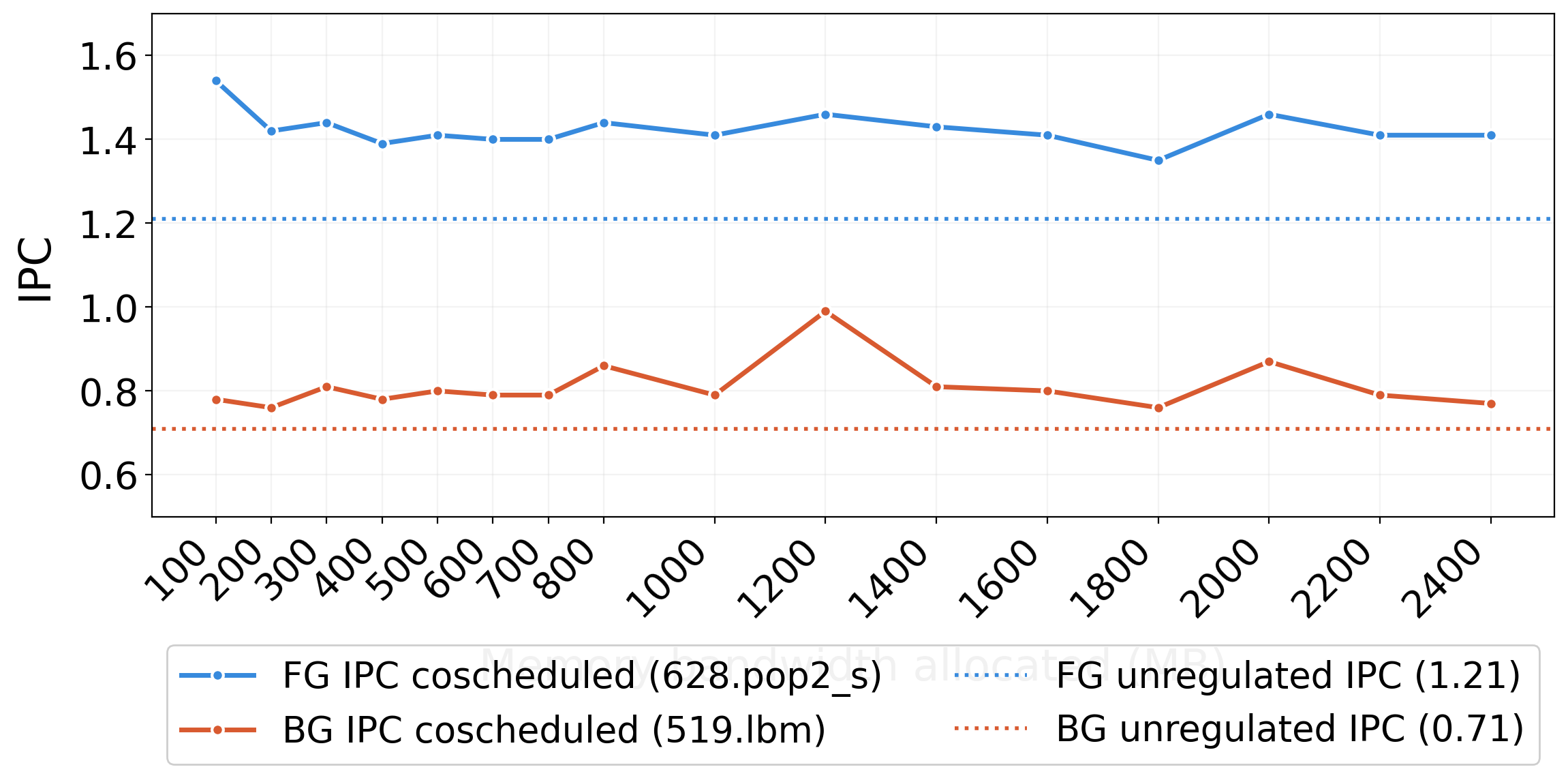}
    \caption{Comparison of IPC performance: Unregulated vs throttled with fixed bandwidth budget. X-axis: allocated memory bandwidth budget (in MB/s), Y-axis: Instructions per cycle}
    \label{fig:unreg_vs_throttle}
\end{figure}

To validate this premise, we co-execute two SPEC CPU 2017 benchmarks—a foreground task (FG) and a more memory-intensive background (BG) task on isolated cores and compare performance using instructions-per-cycle (IPC) measured via \textit{perf} tool, with the BG task showing lower IPC. Without regulation, both workloads suffer an IPC drop (dotted lines, Fig.~\ref{fig:unreg_vs_throttle}). We then allocate each core a fixed bandwidth budget per period T, tracked via performance monitoring counters (PMCs) and replenished each cycle, throttling any core that exhausts its budget. This static allocation yields higher IPC, (solid lines in Fig.~\ref{fig:unreg_vs_throttle}) indicative of better performance. The primary reason for this behavior is that the workloads experience contention at the DRAM. A simple  static allocation and throttling of cores reduces the likelihood of contention at the  DRAM controller.

We make the following contributions in this work:
\begin{itemize}
    \item We propose a novel kernel space memory bandwidth regulation mechanism LMS-AR: LMS Prediction-based Adaptive Regulator, which can periodically estimate the memory bandwidth requirement of the workloads running, by using an adaptive filtering technique.
    \item We implement our proposed LMS-AR scheme outside of the cores as a loadable Linux kernel module and deploy it on COTS hardware.
    \item We execute various types of workloads from SPEC CPU 2017 to evaluate the performance of LMS-AR regulation scheme.
\end{itemize}

\section{System Overview}
\begin{figure}[htbp]
\vspace{-0.1in}
        \centering        \includegraphics[width=1\columnwidth,keepaspectratio]{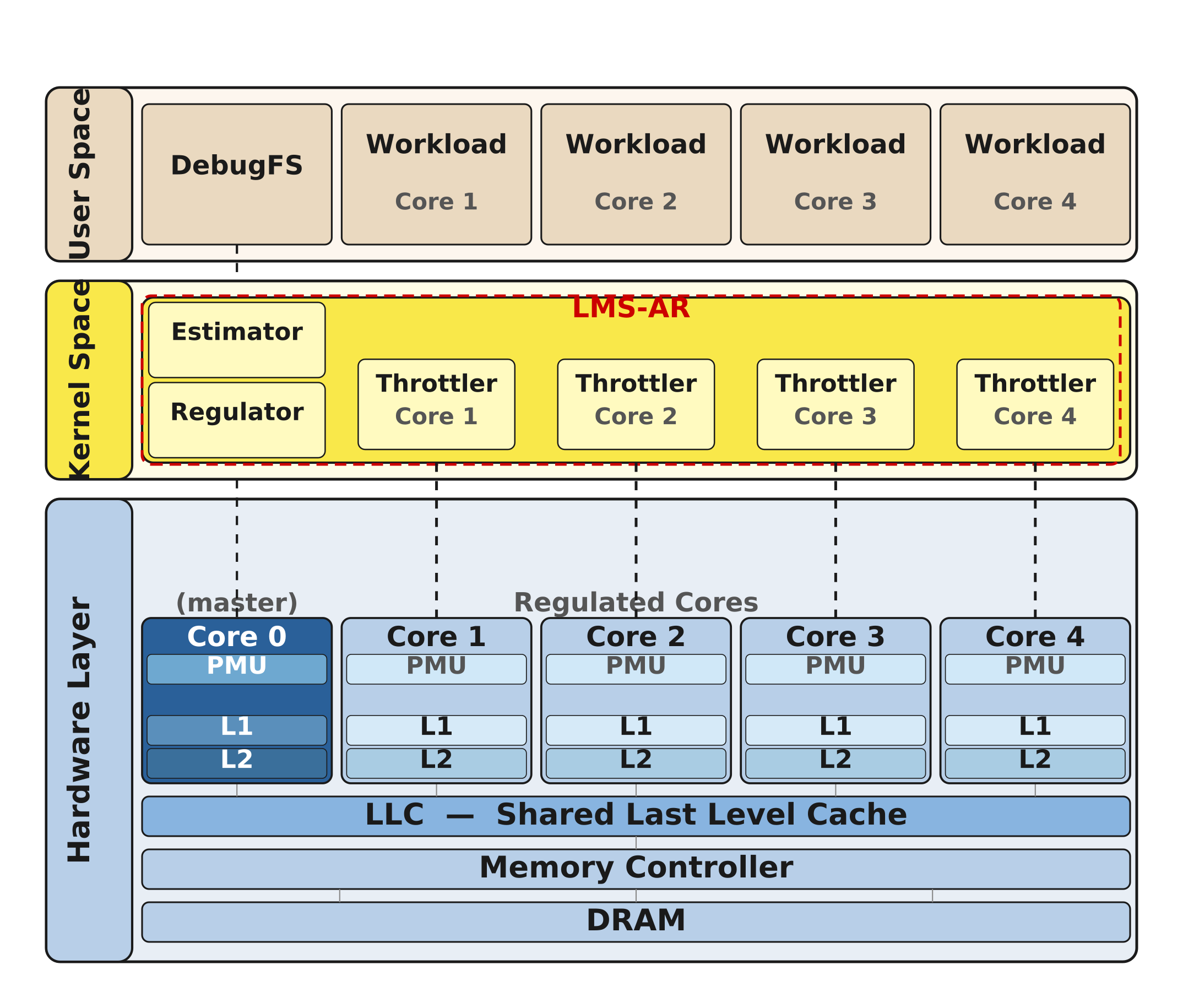}
        \caption{System Overview: LMS-AR's estimator-regulator components executes in kernel space entirely running on Core 0 (Master core)}
        \label{fig:system_overview}
\end{figure}

The primary components of the system architecture are illustrated in Fig.~\ref{fig:system_overview}. The hardware setup consists of a  multi-core platform, which interfaces with a shared DRAM. The platform used supports a PMU for each CPU core, that helps to track the Last Level Cache (LLC) misses and write-backs due to load or store instructions.
The core that runs the regulator is termed as the \textit{master core}  and the ones which are throttled are called the \textit{regulated cores}.

The proposed LMS-AR memory bandwidth regulator resides as a module within the kernel space (shown as dotted rectangle in Fig. ~\ref{fig:system_overview}). It contains a bandwidth estimator and a bandwidth regulator that is exclusively executed on the master Core (Core 0). The regulator does not maintain a timer, however it periodically wakes up to perform estimation and replenishment of the bandwidth for each core. This interval between successive resource replenishments is treated as the \textit{regulation period} $T_{reg}$. The bandwidth estimator analyzes  the memory bandwidth consumed by  each CPU core during the past regulation periods and estimates the requirement for the upcoming regulation interval.
The bandwidth regulator is responsible to 1) replenish the bandwidth to each CPU core using a resource allocation scheme, and 2) \textit{un-throttle} the CPU core at start of the regulation interval if the core is in throttled state. 
The regulator ensures that bandwidth consumed does not exceed a predefined bandwidth allocation for the system as a whole.
Additionally, each core is associated with 1) a high priority task which acts as throttler that suspends the core and thereby blocks execution of tasks in that core, in turn blocking the memory transactions originating from the core, and 2) an interrupt handler which is triggered when the bandwidth budget is exhausted. 

The CPU generates a lot of read and write memory transactions typically known as the load and store instructions. Upon a load or store instruction resulting in a cache miss, a read transaction is issued to retrieve the corresponding cache block from main memory. Should the evicted line carry dirty data, it is flushed to memory through a write transaction known as a cache write-back. The total number of memory accesses incurred by a core can be approximated by summing LLC misses and write-back transactions. Hence, the estimator keeps track of the number of LLC cache misses and write-backs generated by each CPU core that is available from PMU in order to compute the future requirements for bandwidth.

\section{LMS-AR: LMS Prediction-based Adaptive Regulator}

Our proposed LMS-AR regulation mechanism consists of a LMS Prediction-based bandwidth estimator, an adaptive regulator and the necessary logic to throttle the core. These proposed blocks are explained in detail in the following subsections.

\subsection{LMS based Estimator}
\label{subsec:estimator}
The memory bandwidth demands of workloads running on individual cores are inherently 
dynamic and lack discernible patterns that could be reliably exploited for 
workload-level estimation. As a result, the predictor must meet three key 
requirements: (1) it must accurately forecast bandwidth demand for the upcoming 
regulation interval; (2) it must respond quickly to fluctuations in bandwidth 
consumption; and (3) it must be simple enough to implement within a Linux kernel 
module with acceptable runtime overhead.

These requirements make the Least Mean Square (LMS) algorithm a natural fit for the prediction model. The LMS algorithm~\cite{adaptiveProcessing_widrow} 
is a well-established adaptive filtering technique that iteratively refines filter 
coefficients to minimize the mean square error between a desired signal and the 
filter output. Its low computational complexity and ease of implementation make it 
attractive for a wide range of prediction and time-series forecasting tasks.

The prediction model maintains a per-core history of observed memory bandwidth 
and estimates the bandwidth requirement for the next regulation interval as a 
weighted linear combination of past observations. The predictor weights are learned 
online through recursive updates performed at the end of each regulation period. Specifically, we employ the LMS predictor to forecast the memory bandwidth demand of each CPU core based on its consumption over preceding regulation intervals. \\
Let $b_c[t]$ denote the measured bandwidth of core $c$ at interval $t$. The 
predicted bandwidth at interval $(t+1)$ is expressed as:
\begin{equation}
\hat{b}_{c}[t+1] = \sum_{i=0}^{N-1} w_i[t] \cdot x[t-i] = \mathbf{W}[t]^{T} \mathbf{X}[t]
\end{equation}

where $N$ is the number of historical samples retained, and $w_i[t]$ for 
$i = 0, \ldots, N-1$ are the predictor weight coefficients, which are unknown 
at the start of each prediction cycle. The weight vector and input vector are 
defined as:

\begin{equation}
\mathbf{W}[t] = \bigl[w_0,\, w_1,\, \ldots,\, w_{N-1}\bigr]^{T}
\end{equation}

\begin{equation}
\mathbf{X}[t] = \bigl[x[t],\, x[t-1],\, \ldots,\, x[t-N+1]\bigr]^{T}
\end{equation}

The weights are adapted to minimize the prediction error $e[t]$, defined as the 
difference between the true and predicted bandwidth values:

\begin{equation}
e[t] = b_{c}[t+1] - \hat{b}_{c}[t+1]
\end{equation}

Substituting Equation~(1) into Equation~(4) yields:

\begin{equation}
e[t] = b_{c}[t+1] - \mathbf{W}[t]^{T} \mathbf{X}[t]
\end{equation}

The weight vector is initialized to a small value and updated at the end of 
each regulation interval according to the standard LMS update rule:

\begin{equation}
\mathbf{W}[t+1] = \mathbf{W}[t] + \eta \cdot e[t] \cdot \mathbf{X}[t]
\end{equation}

where $\eta$ is the learning rate. To reduce sensitivity to the choice of 
$\eta$, a normalized variant of Equation~(6) is adopted, in which the step 
size is scaled by the energy of the input vector.

\begin{algorithm}[!htbp]
\begin{algorithmic}[1]
\Require Per-core read/write PMU counters, circular history buffers
$H^{r}_{c}$, $H^{w}_{c}$, LMS weight vectors $\mathbf{W}^{r}_{c}$,
$\mathbf{W}^{w}_{c}$, per-core bandwidth limit $B^{max}_{c}$,
global bandwidth pool $B_{pool}$, history window \texttt{HIST\_SIZE}
\Statex
\While{master thread is active}
    \For{each regulated core $c$}
        \Statex \hspace{1.5em} \textbf{Bandwidth Measurement}
        \State Read the PMU counters for core $c$
        \State Determine $B^{r}_{c}$ and $B^{w}_{c}$ as the difference \newline between successive counter readings
        
        \Statex \hspace{1.5em} \textbf{Bandwidth Prediction}
        \State $H^{r}_{c}[i] \leftarrow B^{r}_{c}$
        \State $H^{w}_{c}[i] \leftarrow B^{w}_{c}$
        \State $\hat{B}^{r}_{c} \leftarrow \mathbf{W}^{r\top}_{c} \cdot H^{r}_{c} + B^{init}_{c}$
        \State $\hat{B}^{w}_{c} \leftarrow \mathbf{W}^{w\top}_{c} \cdot H^{w}_{c} + B^{init}_{c}$
        
        \Statex \hspace{1.5em} \textbf{Bandwidth Allocation}
        \State $\hat{B}^{total}_{c} \leftarrow \hat{B}^{r}_{c} + \hat{B}^{w}_{c}$
        \If{$\hat{B}^{total}_{c} \leq B^{max}_{c}$}
            \State $A^{r}_{c} \leftarrow \hat{B}^{r}_{c}$; \quad $A^{w}_{c} \leftarrow \hat{B}^{w}_{c}$
        \Else
            \State $A^{r}_{c} \leftarrow B^{max}_{c} \cdot \dfrac{\hat{B}^{r}_{c}}{\hat{B}^{total}_{c}}$; \quad $A^{w}_{c} \leftarrow B^{max}_{c} \cdot \dfrac{\hat{B}^{w}_{c}}{\hat{B}^{total}_{c}}$
        \EndIf
        \State $B_{pool} \mathrel{+}= \max\!\left(0,\ B^{max}_{c} - (A^{r}_{c} + A^{w}_{c})\right)$
        
        \Statex \hspace{1.5em} \textbf{Counter Replenish}
        \State Program PMU counters $\leftarrow$ $A^{r}_{c}$, $A^{w}_{c}$
        \State Unthrottle core $c$
        \State Restart PMU read and write counters
        
        \Statex \hspace{1.5em} \textbf{LMS Weight Update}
        \State $e^{r}_{c} \leftarrow B^{r}_{c} - \hat{B}^{r}_{c,prev}$
        \State $e^{w}_{c} \leftarrow B^{w}_{c} - \hat{B}^{w}_{c,prev}$
        \State $\mathbf{W}^{r}_{c} \leftarrow \mathbf{W}^{r}_{c} + \eta \cdot e^{r}_{c} \cdot H^{r}_{c}$
        \State $\mathbf{W}^{w}_{c} \leftarrow \mathbf{W}^{w}_{c} + \eta \cdot e^{w}_{c} \cdot H^{w}_{c}$
        
        \Statex \hspace{1.5em} \textbf{Advance History Index}
        \State $i \leftarrow (i + 1) \bmod \texttt{HIST\_SIZE}$
        \State $\hat{B}^{r}_{c,prev} \leftarrow \hat{B}^{r}_{c}$; \quad $\hat{B}^{w}_{c,prev} \leftarrow \hat{B}^{w}_{c}$
    \EndFor
    \State \textbf{sleep} $T_{reg}$
\EndWhile
\end{algorithmic}
\caption{LMS-AR Adaptive Regulation}
\label{Algorithm1}
\end{algorithm}

\subsection{Regulator}
The regulation-estimation scheme is described in Algorithm~\ref{Algorithm1}. At every regulation interval, the master core determines the regulated core's consumed read and write bandwidths $B_c^r$ and $B_c^w$ and appends them to per-core circular history buffers $H_c^r$ and $H_c^w$, respectively. The LMS predictor is then invoked per core and per traffic class(i.e, reads and writes separately), yielding $\hat{B}_c^r$ and $\hat{B}_c^w$ from the weight vectors $\mathbf{W}_c^r$ and $\mathbf{W}_c^w$. A small statically-chosen per-core bias $B^{init}_{c}$ is added to the LMS output to guard against under-prediction during the cold-start period before weights have converged.If $\hat{B}_c^{\mathrm{total}} = \hat{B}_c^r + \hat{B}_c^w$ fits within the per-core ceiling $B_c^{\max}$, the full predicted amounts are allocated; otherwise, the ceiling is split proportionally, with any surplus returned to a global pool $B_{\mathrm{pool}}$ available to other cores. The allocations are programmed into the PMU period registers, counters are restarted, and the core is un-throttled. At the end of the interval, per-class prediction errors update $\mathbf{W}_c^r$ and $\mathbf{W}_c^w$ via the LMS rule described in Section ~\ref{subsec:estimator}.

\subsection{Stalling the core}
The functionality of stalling the core is an important step in preventing the core from consuming more memory bandwidth than what is allocated. 
In this work, a throttler thread similar to \cite{memguard2013} is used and certain architecture specific instructions (such as MWAIT and MONITOR for x86) are executed indicating the core to suspend execution, which may hint the processor to move to low power states supported by the OS. The actual transition to and from low power state is dependent on the firmware configuration.

\subsection{Using FPU within Linux kernel}
Estimating memory bandwidth requires floating-point arithmetic — specifically products and sums over measured values — to achieve a reasonably accurate result. However, floating-point computation is generally discouraged in kernel modules, as it introduces non-trivial overhead and risks unsafe handling of the Floating Point Unit (FPU) context. 
Here, we explicitly enable the floating-point unit within kernel space, carefully bracketing only the necessary calculations (Lines 21-22 Algorithm~\ref{Algorithm1} ), within floating-point critical sections, in order to keep the implementation safe and lightweight while producing bandwidth estimates that closely track actual consumption. We employ generic runtime APIs (architecture specific implementations) provided in Linux kernel \cite{linux_kernel_FPU} to create these critical sections. These  APIs typically create critical sections by ensuring 1) the preemption is disabled; 2) sanity checks the context i.e, if FPU is usable at the point; 3)  saves the interrupted task's FPU state, if necessary and 4) initializes the control registers to a known value, in that order. The critical sections are free of blocking code or nested calls and use batched floating point computations.

These computations are exclusive to the FPU of the master core - one of the direct advantages of performing out-of-core regulation.

\section{Experimental Evaluations}
\subsection{Metrics}
The IPC slowdown ratio (runalone IPC/ co-scheduled IPC)\cite{memguard2013}, which indicates the extent to which co-scheduling multiple workloads on a shared memory system can degrade individual workload performance relative to running alone, is used as the primary metric to compare the performance of the regulators. A ratio of 1.0 indicates no degradation, while higher values reflect greater interference caused by contention on the shared memory bus.

\subsection{Experimental Setup}
The experiments were conducted on a desktop platform equipped with an Intel® Core™ i5-11400 processor (Rocket Lake), featuring 6 cores and 12 threads with a base frequency of 2.60 GHz. Though the system supports four DIMM slots, for our work it is equipped with a single DRAM module 3200 MT/s. For all the experiments we use 16 GB of dual-channel DDR4-3200 memory, providing a peak memory bandwidth of 50 GB/s. The processor supports hardware-accelerated floating-point operations. For the purpose of this work hyper-threading was disabled in the firmware to ensure the cores being used are actual physical cores. Hardware pre-fetcher was disabled using model specific registers\cite{IntelSDM-MSR} in order to avoid hiding the inherent memory latency between the CPU and DRAM. 
We use the stock linux kernel version 6.8.0 that comes along with Ubuntu 22.04 LTS. \textit{cpusets}\cite{cpusets_user} infrastructure in Linux is used to isolate the CPU cores and selectively execute the desired SPEC CPU 2017 benchmark as a workload. Additionally we use a regulation interval of 1 ms for all our experiments (Refer Section IV-G for explanation).

Though the work demonstrates the results on a 6 CPU core based hardware platform, the software design of the kernel module itself can be scaled to multiple CPU cores with minimal changes to the \href{https://github.com/ss22ongithub/AdaptiveRegulator/tree/main}{source code}.
\subsection{Sustainable Bandwidth Assessment}
Sustainable bandwidth is the bandwidth the memory controller of a DRAM withstands without resulting in over saturation\cite{Zuepke2024},\cite{ewarp} and its assessment is a significant step in setting the limits of how much memory bandwidth a core or a set of cores can utilize in the system without leading to interference. In this context, we reuse publicly available  \textit{bench} tool \cite{Zuepke2024} to perform,  sustainable bandwidth measurements for our platform. Based on the measurements, we fix the sustainable bandwidth at
$B_{sustain}^r =$ 6000 MB/s for reads and $B_{sustain}^w =$ 4000 MB/s for writes on our setup. The details of the assessment can be found in the Technical Report for LMS-AR \cite{LMSAR_Technical_Report}.

\subsection{Workload Profiling}
We compute the average bandwidth consumed for each workload to gain insights about the characteristic of the workload - whether they are read heavy or write heavy- using the \textit{perf} tool. We additionally classify the workloads as High, Medium or Low to indicate its memory bandwidth usage intensity, which is described in the Technical Report for LMS-AR\cite{LMSAR_Technical_Report}.

% as shown in Table~\ref{tab:bw_classification}.

% Define row colors
\definecolor{highbw}{RGB}{255, 204, 204}    % Light red for High
\definecolor{medbw}{RGB}{255, 243, 204}     % Light yellow for Medium
\definecolor{lowbw}{RGB}{204, 238, 204}     % Light green for Low

\subsection{Evaluations on 2-core setup}

\begin{figure*}[!htbp]
    \centering
    \includegraphics[width=0.80\textwidth]{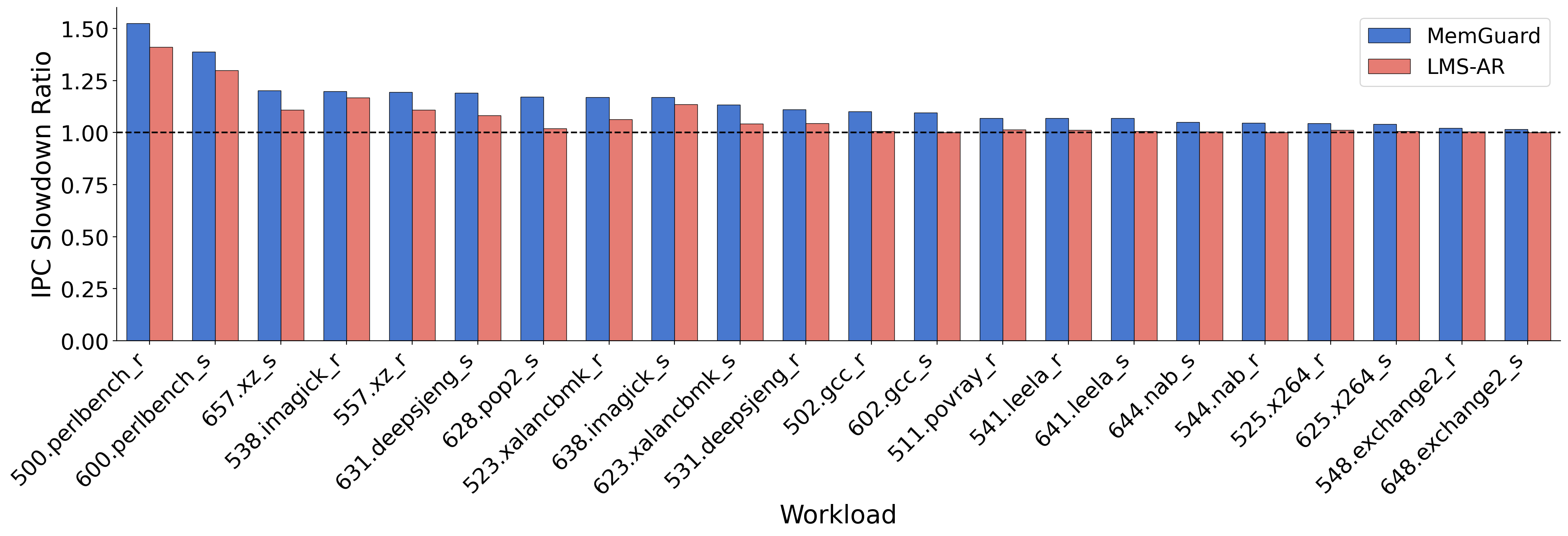}
    \caption{Comparison of Slowdown ratios for system with 2 cores + 1 (master core)}
    \label{fig:2_core_performance}
\end{figure*}

We create a setup where the estimation and regulation mechanism is executed on Master core (core 0) and the regulated cores are  Core 1 and Core 3, designated as C1 and C3 respectively. The Cores C1 and C3 are isolated using the \textit{cpusets} facility provided in the Linux Kernel.
The regulated cores are assigned a bandwidth limit of 1GB/s each such that the total utilized bandwidth does not exceed the overall bandwidth allocated for the system.\par
We select 519.lbm as the background task, for all our experiments, due to its high memory intensity and ability to generate a sustained stream of read and write memory traffic, making it an ideal background workload for stressing the shared memory. The measurements are taken for fixed amount of time (20s) for each set of co-executed tasks.\par
Fig~\ref{fig:2_core_performance} compares the IPC slowdown ratios of MemGuard and the LMS-AR across 22 of the 43 workloads of the SPEC CPU 2017 benchmarks. The improvements are observed across medium and low memory intensive workloads. Across the 22 SPEC CPU 2017 benchmarks in the 2-core setup, LMS-AR achieves an average IPC slowdown ratio of 1.073 with a standard deviation of 0.085, compared to MemGuard's average of 1.133 with a standard deviation of 0.122.
For rest of the workloads (which are predominantly memory-intensive), slowdown ratios of $<$ 1 are observed with LMS-AR as the high LLC misses seen in stand-alone measurements, are regulated efficiently by LMS-AR leading to higher IPC than stand-alone IPC. This behavior is further corroborated by the Table in Section E of the Technical Report on LMS-AR\cite{LMSAR_Technical_Report}.
%Low-contention workloads are barely impacted by either approach (both near 1.0). These benchmarks likely have low memory bandwidth demands, so neither regulation scheme influences their performance.

\subsection{Evaluations on 4-core setup}

\begin{figure}[b]
    \centering
    \includegraphics[width=0.65\columnwidth]{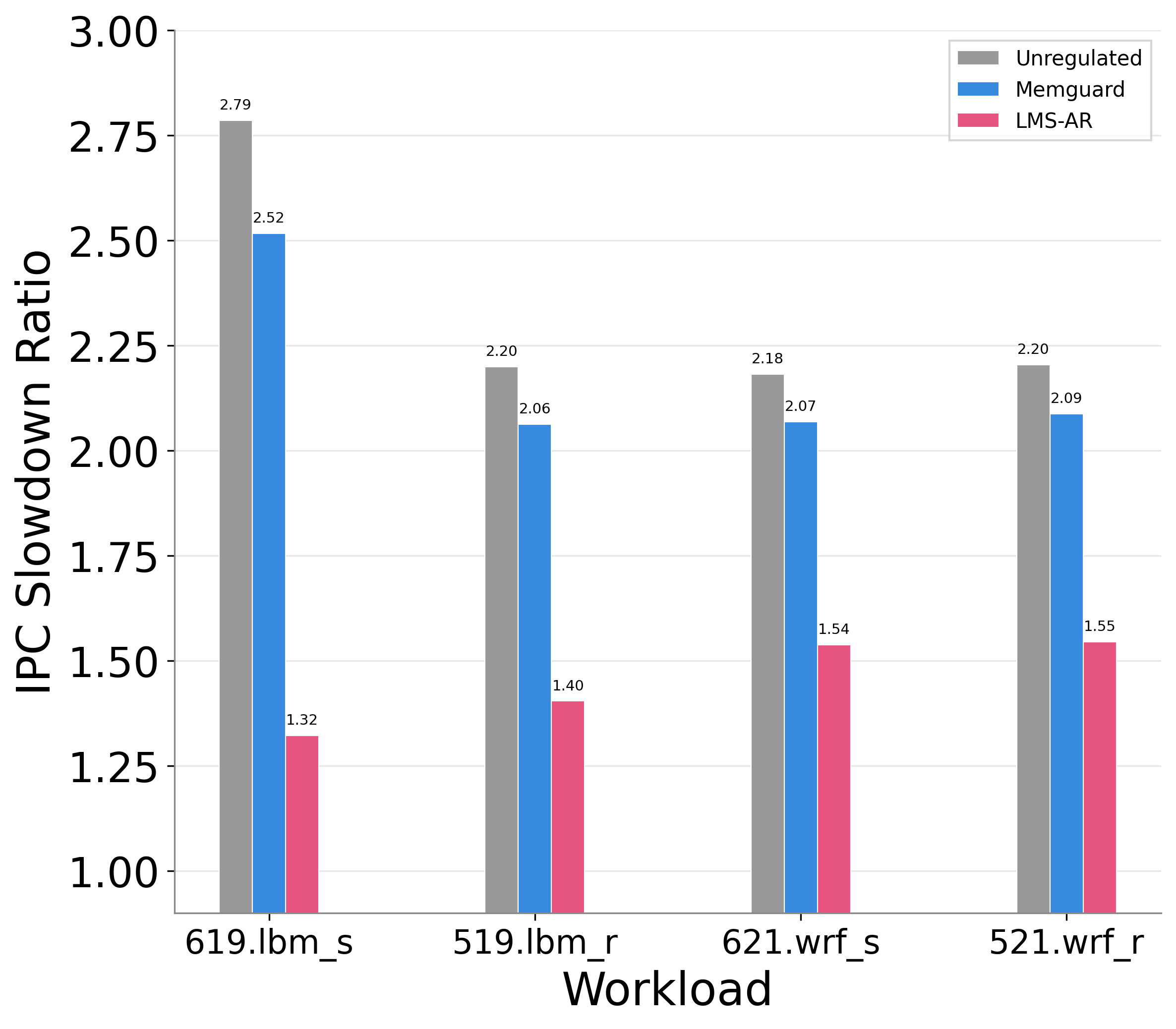}
    \caption{Workload Performance 4 Cores + 1 master core setup with high memory intensive workloads }
    \label{fig:5b_4C_all_H}
\end{figure}

In this section, we compare three scenarios- no regulation, MemGuard regulation, and LMS-AR Regulation — across few of SPEC CPU 2017 workloads in a four-core co-scheduled setup. 
In order to better comprehend the interference effects and minimize the effects of cache on performance, in the first evaluation (Fig~\ref{fig:5b_4C_all_H}) we choose workloads such that memory bandwidth required by the workload exceeds the last level cache size of the platform. The workloads considered here -- 619.lbm\_s,519.lbm\_r,621.wrf\_s,521.wrf\_r -- are classified as highly memory intensive. We consistently observe this behavior of LMS-AR regulating efficiently with other memory intensive workloads as well. This aligns with our initial premise that an adaptive distribution of the memory bandwidth leads to lower interference.\par

\begin{figure}[!htbp]
    \centering
    \includegraphics[width=0.65\columnwidth]{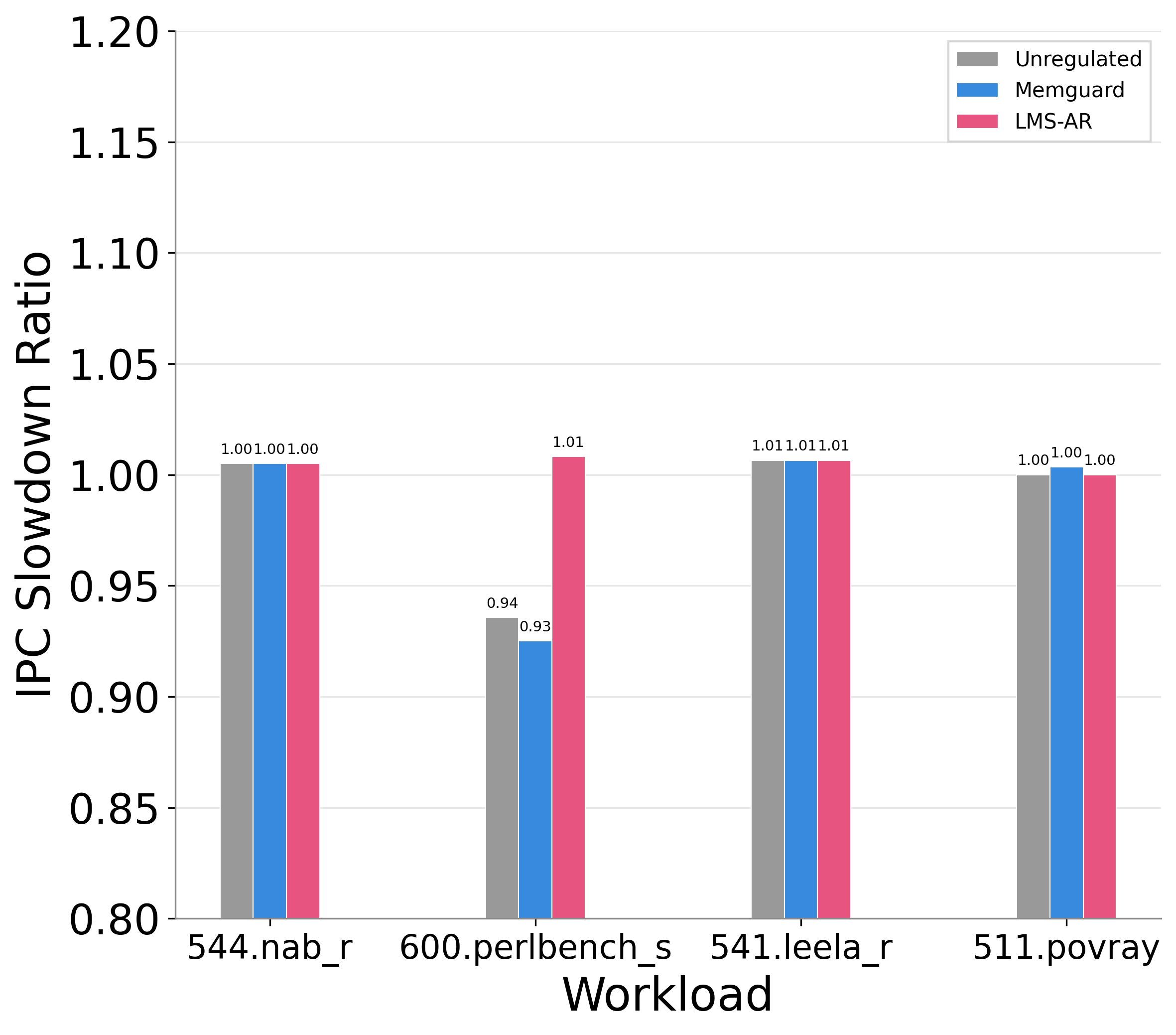}
    \caption{Workload Performance 4 Cores + 1 master core setup with low memory intensive workloads}
    \label{fig:4_core_all_L}
\end{figure}

\begin{figure}[!htbp]
    \centering
    \includegraphics[width=0.65\columnwidth]{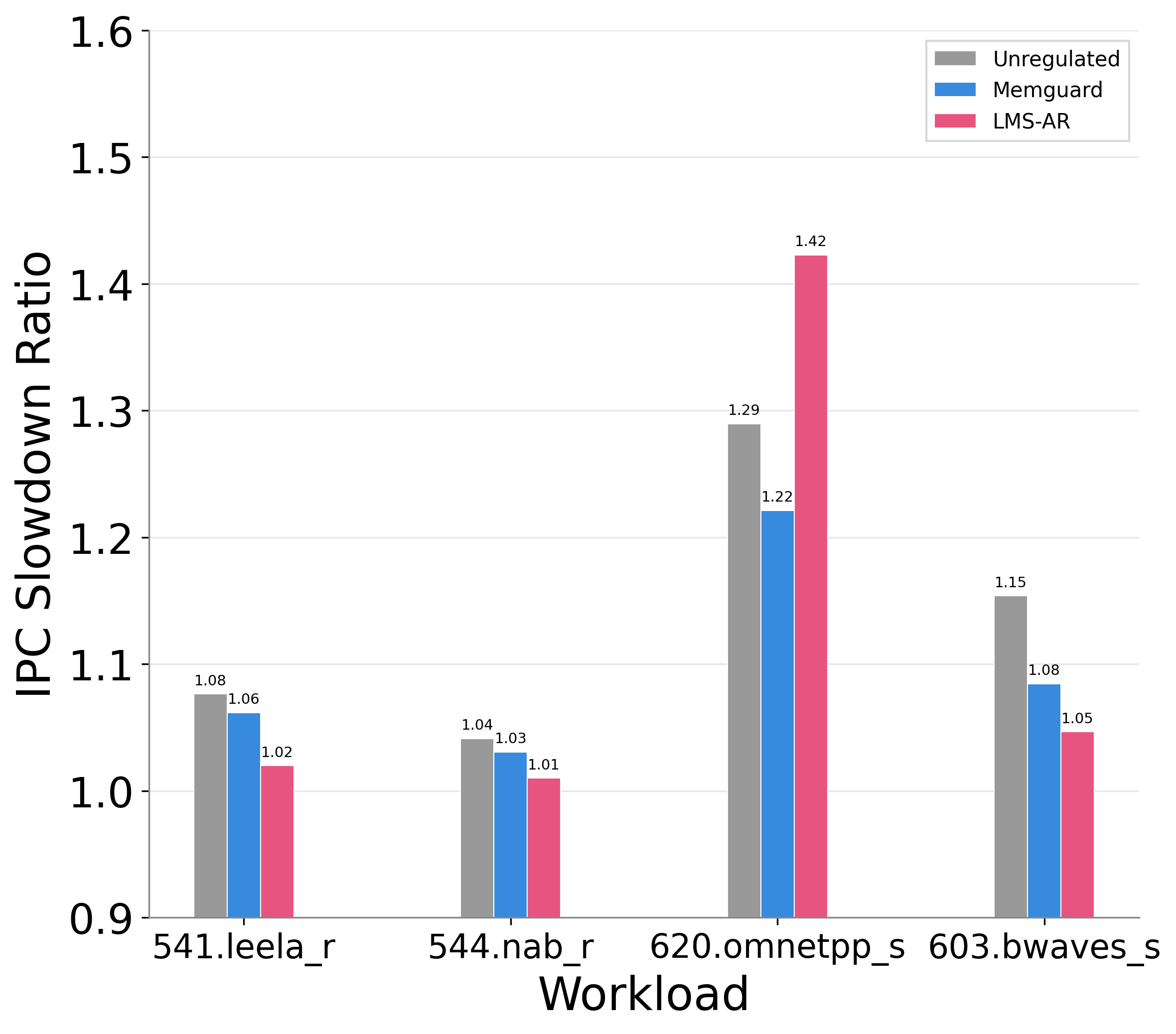}
    \caption{Workload Performance 4 Cores + 1 master core setup with diverse set of workloads }
    \label{fig:5b_4C_diverse}
\end{figure}

In the second experiment, we choose workloads classified as low memory intensity for co-execution. In Fig.~\ref{fig:4_core_all_L}, clearly the regulation mechanism has little impact on the performance as workloads tend to operate very close to the sustainable bandwidth (slowdown ratio close to 1) and the allocations satisfy their smaller bandwidth requirement.\par
In a diversified workload setting (Fig~\ref{fig:5b_4C_diverse}) with -- 603.bwaves\_s(H), 541.leela\_r(L), 544.nab\_r(L) and 620.omnetpp\_s(M), we find that LMS performs worse for medium intensity workload, attributed to weights converging on a regulation budget that is too restrictive for this workload's bursty memory behavior.
There is limited observation from this work that a computationally involved  model (such as the LMS) for predicting bandwidth requirement could produce better estimates, which helps to allocate the right amount of bandwidth during each regulation period and improves over time. Additionally, as the core count increases in the system, we expect the workloads will continue to benefit from this solution. 
%However, the impact of last level cache on the high-intensity workloads, which require more detailed study, with larger number of contenders, is left for future work.

\subsection{Impact of Regulation Interval}
The regulation interval defines the fixed temporal period between successive invocations of the bandwidth control loop. Within each interval, cores execute and consume bandwidth according to their allocated budgets. Upon interval completion, actual bandwidth consumption is measured and new allocations are computed for the next interval. In our implementation, a master thread wakes periodically at fixed intervals to perform these measurements and allocation operations.

\begin{figure}
    \centering
    \includegraphics[width=0.8\linewidth]{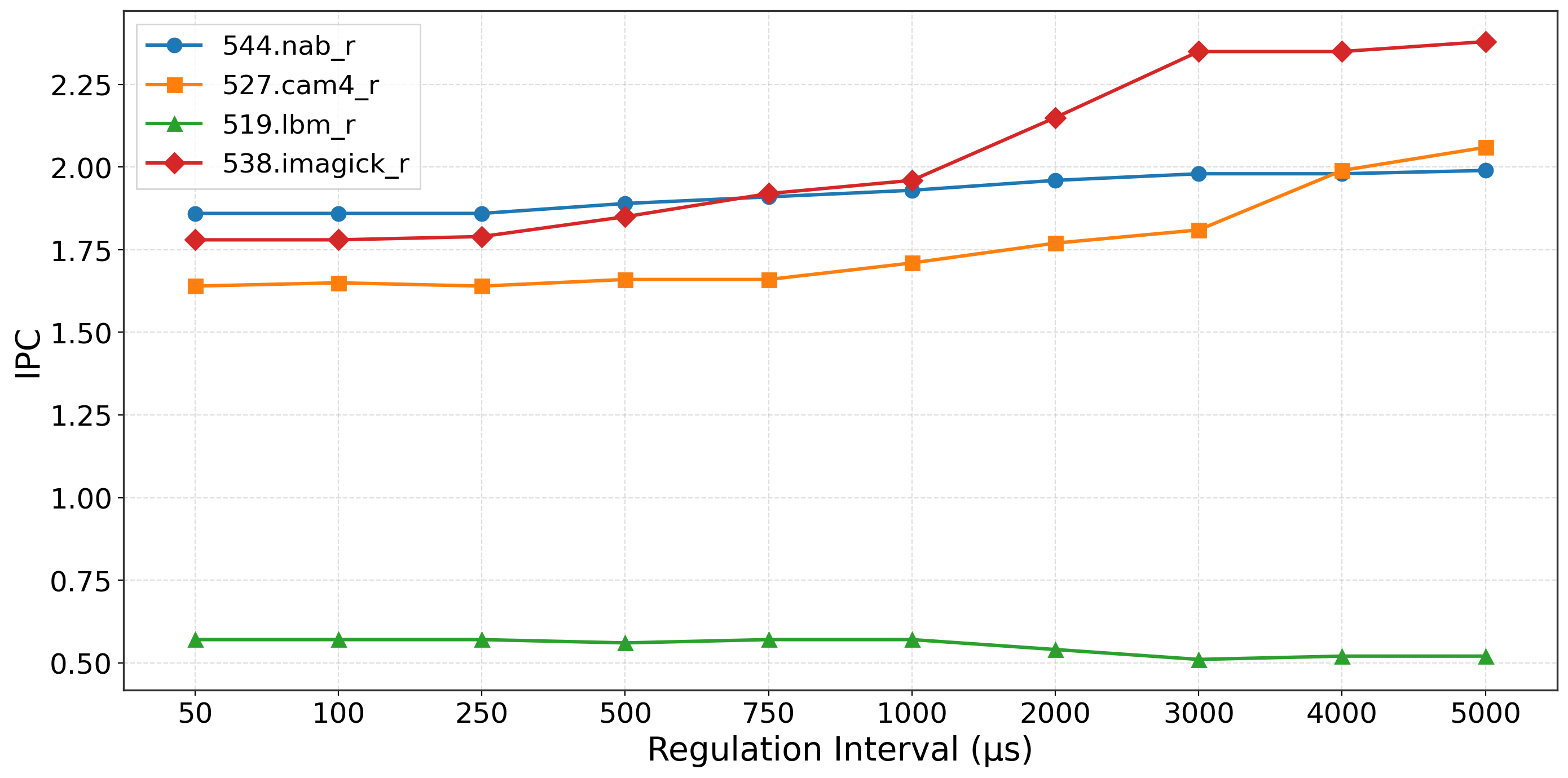}
    \caption{ Variation in IPC performance of  workloads in 4 + 1 core setup with varying regulation intervals (50us to 5ms). }
    \label{interval_vs_ipc}
\end{figure}

The regulation interval determines effectiveness of regulation mechanism, representing a trade-off between responsiveness and overhead which involves finding the regulation interval value that requires balancing two competing concerns. A long regulation interval can lead to poor estimates or delayed adaptation, leading to under allocation. A very small regulation interval can lead to excessive overhead and insufficient observation time, that captures random fluctuations instead of real workload characteristics or pattern.\par

The variation in IPC values for each workload, when co-scheduled on the 4-core setup, is illustrated in Fig.~\ref{interval_vs_ipc}. Lower memory-intensive workloads (538.imagick and  544.nab) benefit most with increasing regulation interval, as longer regulation interval allows for a longer execution period. Workloads with higher memory requirement (such as 519.lbm) essentially maintain the same performance. At lower (less than $500{\mu}s$) intervals, the advantage diminishes as frequent intervention in the control loop leads to overheads and causes the prediction mechanism to underperform.

\section{Related Work}
The problem of temporal isolation with main memory as the key contributor for interference in mixed critical systems has been studied and analyzed extensively\cite{Lugo2022},\cite{Zini_Etal_2022}.
The solutions available in this space range between hardware design solution to software assisted solutions.\par
The hardware-assisted solutions include 1) enforcement hardware \cite{BRU_Farshchi_et_all, memcore_Izhbirdeev_2024}, which virtually eliminates the overhead limits of interrupt-based throttling to achieve very fine-timescale bandwidth accounting, 
and 2) memory-controller hardware designs \cite{duomc_Mirosanlou_2022,slack_based_tdm,drambulism_2020, rtcontroller_Li_2016, cmd_level_prio_sched_Kim_2015} that provide flexible priority-based controller designs with improvement on TDM based designs for performance and predictability 3) vendor-driven
hardware QoS mechanisms Arm MPAM \cite{mpam_zini_2024}

Software-assisted DRAM regulation mostly involves source-throttling mechanisms implemented in the OS or hypervisor, where each core/VM is given a memory budget over a fixed period and is paused when it exceeds that budget. The foundational
work MemGuard and follow-up work in \cite{memguard2013,memgaurd_followup_2016} have been integrated into schedulability and mixed-criticality analysis \cite{sched_1_AWAN_2028},\cite{sched_2_yao_2016}.
E-Warp\cite{ewarp} uses \cite{memguard2013} to regulate the CPU core, but also extends to regulating traffic from an accelerator.  \textit{MemPol} \cite{Zuepke2024}, on the other hand, uses an out-of-core approach to regulate the cores, to reduce the memory transactions generated by per-core interrupt handlers in \textit{MemGuard},
enabling microsecond-scale on–off control with sliding-window policing and redistribution of unused bandwidth without disturbing the application cores with frequent interrupts, were also one of the motivations in our work. The work in \cite{MemUtil} shifts monitoring closer to the true DRAM bottleneck by using controller-provided utilization counters inside a feedback loop, rather than relying solely on PMU. However this is applicable to specific hardware platforms that can support profiling statistics of the DRAM controller.

LMS-AR  closely relates to \textit{MemGuard} in that both are 1) Linux kernel modules that leverage hardware PMCs to regulate shared resource usage 2) share the design philosophy of decomposing resource allocation into a guaranteed component for real-time tasks and a best-effort component for non-critical workloads 3) employs reclaiming mechanisms to minimize resource wastage from static partitioning. However, the two approaches differ significantly in their prediction and regulation strategies. LMS-AR incorporates a Least Mean Squares (LMS) adaptive filtering technique for prediction, to minimize the error between predicted and observed resource demand in real time, enabling more accurate tracking of workload and faster convergence.
We do not compare with \textit{MemPol}, though our work has similarities(an out-of-core design), as the master core that regulates other cores shares the same DRAM as the regulated cores and the platform lacks vendor-specific debug interface. However, the design enables exploring sophisticated prediction models with little impact on regulation performance.

\section{Conclusion}
In this paper, we presented LMS-AR, a Linux kernel module for adaptive resource regulation on multicore platforms. It employs model-based control with real-time performance monitoring to dynamically adjust resource allocation based on workload characteristics. Experimental results demonstrate that LMS-AR improves average-case performance for real-time tasks co-executing with best-effort workloads. These have shown to improve the performance of most of the low, medium and high intensity workloads. However, we will explore the efficiency of learning based prediction models in future work along with extending the framework to support heterogeneous computing platforms.
\bibliographystyle{IEEEtran}
\bibliography{ref}

\end{document}